\documentclass[preprint]{elsarticle}

\usepackage{booktabs} % For formal tables
\usepackage{subcaption}
\usepackage{multirow}
\usepackage{amsmath}
\usepackage{fancyvrb}
\usepackage{color}

\usepackage[margin=1in]{geometry}

\usepackage{paralist, tabularx}

\usepackage{hyperref}

\journal{Information Processing and Management}

\begin{document}

\begin{frontmatter}

\title{Utilizing Microblogs for Assisting Post-Disaster Relief Operations via Matching Resource Needs and Availabilities\tnoteref{t1}}

\tnotetext[t1]{\textcolor{blue}{{\bf This work has been published in Information Processing and Management, Elsevier, vol. 56, issue 5, pp.~1680--1697, September 2019.}}}

\author[iitkgp]{Ritam Dutt\corref{eqcon}}
\ead{ritam.dutt@gmail.com}
\author[iiest,uem]{Moumita Basu\corref{eqcon}}
\ead{moumita.basu@uem.edu.in}
\cortext[eqcon]{The first two authors contributed equally to the work.}
\author[iitk]{Kripabandhu Ghosh}
\ead{kripa.ghosh@gmail.com}
\author[iitkgp]{Saptarshi Ghosh}
\ead{saptarshi@cse.iitkgp.ac.in}

\address[iitkgp]{Indian Institute of Technology, Kharagpur, India}
\address[iiest]{Indian Institute of Engineering Science and Technology, Shibpur}
\address[uem]{University of Engineering and Management, Kolkata}
\address[iitk]{Indian Institute of Technology, Kanpur, India}

\begin{abstract}
During a disaster event, two types of information that are especially useful for coordinating relief operations are needs and availabilities of resources (e.g., food, water, medicines) in the affected region.
Information posted on microblogging sites is increasingly being used for assisting post-disaster relief operations.
In this context, two practical challenges are (i)~to identify tweets that inform about resource needs and availabilities (termed as need-tweets and availability-tweets respectively), 
and (ii)~to automatically match needs with appropriate availabilities.
While several works have addressed the first problem, there has been little work on automatically matching needs with availabilities.
The few prior works that attempted matching only considered the resources, and no attempt has been made to understand other aspects of needs/availabilities that are essential for matching in practice.
In this work, we develop a methodology for understanding five important aspects of need-tweets and availability-tweets, including what resource and what quantity is needed/available, the geographical location of the need/availability, and who needs / is providing the resource. 
Understanding these aspects helps us to address the need-availability matching problem considering not only the resources, but also other factors such as the geographical proximity between the need and the availability. 
To the best of our knowledge, this study is the first attempt to develop methods for understanding the semantics of need-tweets and availability-tweets. 
We also develop a novel methodology for matching need-tweets with availability-tweets, considering both resource similarity and geographical proximity.
Experiments on two datasets corresponding to two disaster events, demonstrate that our proposed methods perform substantially better matching than those in prior works. Additionally, our proposed methodologies are reusable across different types of disaster events.
\end{abstract}

\begin{keyword}
Disaster\sep microblogs \sep post-disaster relief\sep resource needs\sep resource availabilities\sep matching needs and availabilities
%{\MSC[2010] 00-01\sep 99-00}
\end{keyword}

\end{frontmatter}

%\linenumbers

\section{Introduction} \label{sec:intro}

\noindent
During a disaster situation (e.g. floods, earthquakes, hurricanes), microblogging sites like Twitter and Weibo are very useful for gathering
situational information~\cite{social-media-emergency-survey,varga-help-tweets,rudra-cikm-disaster}. 
Two types of information that are especially useful for coordinating post-disaster relief operations are 
{\it what resources are needed} and  {\it what resources are available} in the disaster-affected area.
Specifically, two types of tweets (microblogs) are considered useful by relief workers~\cite{BasuASONAM17,Khosla2017,purohitFM} --
\noindent {\bf (i)~Need-tweets}, which inform about the need or requirement
of specific resources such as food, water, medical aid, shelter, etc., and
%An example of Need-tweet posted during the 2015 Nepal earthquake is -- \textit{Nepalis, r w/o water \& electricity. Water is essential to be supplied to the affected people in Nepal.} 
\noindent {\bf (ii)~Availability-tweets}, which inform about the availability
of specific resources in the affected region, or potential availability, e.g., resources being transported
to the region. 
%An example of availability-tweet is --  \textit{@UPGovt sends 21 trucks of mineral water, biscuits and medicines to \#earthquake affected \#NepalQuake}.
%Note that, a tweet can be both a need-tweet and an availability-tweet if it mentions about need for one resource and availability of another resource.

For coordinating relief operations, two tasks are necessary to be automated~\cite{Khosla2017,purohitFM} --
 {\bf (1) Identifying need-tweets and availability-tweets} 
from among thousands of tweets posted during a disaster situation, most of which contain conversational content~\cite{social-media-emergency-survey}, and
{\bf (2) Matching need-tweets with appropriate availability-tweets}, which would help relief workers to satisfy the needs quickly.
The first problem of identifying need-tweets and availability-tweets has been addressed in several prior works~\cite{BasuASONAM17,Khosla2017,purohitFM,varga-help-tweets} (see Related Work section for details).
However, the second problem of matching need-tweets with suitable availability-tweets has received relatively less attention.
We focus on this matching problem in this work. 
We assume that a set of need-tweets and a set of availability-tweets have already been identified from a given (large) set of tweets posted during a disaster event,
and we focus on the matching of the identified needs with appropriate resource availabilities.

Table~\ref{tab:need-available-examples}
shows some examples of need-tweets and matching availability-tweets, from one of the datasets used in this work, that were retrieved by one of the methodologies proposed in this work (datasets and methodologies detailed in subsequent sections).
While some pairs might be easier to match since they state exactly the same resource (e.g., `water' or `medical supplies'), some
pairs are much more difficult to match because they use different terms to refer to the same resource -- e.g., need of `shelter' and availability of `tents', need of `bread and roof' and availability of `shelter' and `food' -- as shown in Table~\ref{tab:need-available-examples}.

\begin{table}[tb]
 \footnotesize
 \centering
 \begin{tabular}{|p{0.45\textwidth} | p{0.45\textwidth} |}
 \hline
  \textbf{Need-tweet (excerpts)} & \textbf{Availability-tweet (excerpts)}\\
  \hline
  Mobile phones are not working, no electricity, no \textbf{water} in \#Thamel, \#Nepalquake  &   Please contact for drinking free service \textbf{water} specially for Earthquake Victim. Sanjay Limbu [mobile num]
\\ \hline
Over 1400 killed.  Many Trapped. \textbf{Medical Supplies} Requested.   & 20,000 RSS personnel with \textbf{medical supplies} and other help the first to reach earthquake damaged zones in \#Nepal
\\  \hline
Nepal earthquake: thousands in need of \textbf{shelter}  in country little able to cope [url]   &   can anyone we know pick the 2000 second hand \textbf{tents} from Sunauli and distribute it to the people in need in Nepal?  
\\ \hline
Victims of \#Earthquake in Nepal are waiting your support for \textbf{bread and roof}. Thousands of people are still in open space & \#AOL Nepal Centre is converted into a rescue \textbf{shelter}, sheltering 100's of ppl where volunteers r providing \textbf{food}, water
\\ \hline
\textbf{Power} Outages Plague Nepal: Darkness descended on much of Nepal after a powerful earthquake rattled the country [url]& @thehimalayan: NEA resumes supplying \textbf{electricity} to Nepal [url]\\
\hline
\end{tabular}
\caption{{\bf Examples of need-tweets and matching availability-tweets (from a dataset of tweets on 2015 Nepal earthquake) that were identified by one of the methodologies proposed in this work (detailed in subsequent sections). The common resources for each pair are shown in boldface.}}
\label{tab:need-available-examples}
\end{table}

\subsection{Objectives of the study} \label{subsec:objective}

The objective of the present study is to assist post-disaster relief operations, by developing a practical  methodology for matching need-tweets and availability-tweets (resource needs and availabilities extracted from microblogs).
To this end, we utilize different factors associated with need-tweets and availability-tweets, such as similarity of the resources mentioned, and location proximity, to perform the matching more effectively than what prior works attempted (as explained below).
%Our overall motivation is to provide an efficient methodology to  assist disaster management authorities towards solving the practical spatio-temporal resource optimization problem.

%\noindent\textcolor{blue}{ \bf Prior works}
\subsection{Prior works and their limitations}\label{subsec:prior-work}

To our knowledge, only two prior works have attempted to address the matching problem~\cite{purohitFM,matching-www18-poster}.
Purohit {\it et al.}~\cite{purohitFM} attempted to match based on the resources mentioned in the tweets, while
our previous work~\cite{matching-www18-poster} performed the matching based on either the resources mentioned or the full text of the tweets (details in Related Work section). 
Thus, both these prior works focused only on the resources mentioned in tweets.
Neither of the prior works attempted to identify other factors that should be considered in practice for matching needs and availabilities in a post-disaster scenario, e.g., the quantity needed or available, the specific geographical location of the need / availability, and so on.
For instance, to match the first need-tweet shown in Table~\ref{tab:need-available-examples}, it needs to be considered that the need for water is in `Thamel' (a town in Nepal), and hence availabilities near this location should be sought.
In other words, the prior works addressed very simplified versions of the need-availability matching problem.
The present work takes a significant step towards solving the actual problem of matching resource needs and availabilities, by identifying the different aspects of resource-needs and availabilities, such as 
the specific resources which are needed/available, the location of the need/availability, and so on.
%To our knowledge, there has not been any attempt to match post-disaster resource needs and availabilities considering these factors.

\subsection{Contributions of present work}\label{subsec:contribution}

The contributions of the present work are as follows.
\begin{itemize}
 \item
 %In this work, we take the first step towards identifying the different factors that need to be considered for the post-disaster resource matching problem. 
We develop a novel methodology for understanding the semantics of need-tweets and availability-tweets, and extract five specific types of information from such tweets --
(i)~the resource that is needed/available,
(ii)~the quantity that is needed or available,
(iii)~the geographical location where the resource is needed or available,
(iv)~the source who is offering the resource, and
(v)~contact information (phone number, email, etc.) of the source or the ones needing the resource.
To our knowledge, no attempt has been made to understand the different aspects of resource-needs and availabilities from the tweets (apart from the resources mentioned).
Identifying these information enables us to perform the matching considering not only the resource(s) but also other factors such as the geographical distance between the need and the availability.

\item We perform extensive experiments on need-tweets and availability-tweets posted during two recent earthquake events - the Nepal earthquake in 2015, and the 2016 earthquake in central Italy (details of datasets are given in Section~\ref{sec:dataset}).
Our experiments show that our proposed methodology can extract the five types of information with high precision and recall.

\item Using the extracted information, we propose two methodologies for matching needs with availabilities, one considering only resources (as done by the prior works) and another considering both resources and geographical proximity of the need and the availability.
%We demonstrate that our proposed methods achieve considerably better need-availability matching, compared to the methods in the prior works~\cite{purohitFM,matching-www18-poster}.
Our methods for matching based only on resources achieve precision more than $0.8$ and recall more than $0.95$, and substantially outperform similar methods in prior works~\cite{purohitFM,matching-www18-poster}.

\item We propose the first methodology (to our knowledge) for matching post-disaster needs and availabilities considering both the resources and the geographical proximity of the need/availability. 
For more than 90\% of need-tweets that mention a location, this methodology can identify availabilities that are geographically close to the location of the need.

\item To investigate the reusability of our proposed methods to other types of  disaster events, we apply our proposed methodologies to tweets posted during another disaster event - floods in the Indian city of Chennai in December 2015. 
We see that our methods can effectively match resource-needs with availabilities for different types of disaster events.
\if 0
\textcolor{blue}{To investigate the reusability of our proposed methods to other types of  disaster events, we apply our proposed methodologies to tweets posted during another disaster event - floods in the Indian city of Chennai in December 2015. 
We see that our methods can effectively match resource-needs with availabilities for different types of disaster events.}
\fi
\end{itemize}

We believe that the methodologies developed in the present work are important in practice for coordinating post-disaster relief operations. Especially, since the proposed methods are unsupervised, they can be readily deployed during an ongoing disaster event.

The rest of the paper is organized as follows. We discuss the Related Work in Section~\ref{sec:related}. Section~\ref{sec:dataset} describes the datasets used for our experiments. We attempt to understand the semantics of need-tweets and availability-tweets in Section~\ref{sec:understanding}. 
Various matching methodologies are discussed in Section~\ref{sec:matching-methods}, and the methods are evaluated in Section~\ref{sec:evaluation}. 
Section~\ref{sec:reuse} explores the reusability of our proposed matching methods on a new disaster event.
The study is  concluded in Section~\ref{sec:conclusion}.
%%%%%%%%%%%%%%%%%%%%%%%%%% SECTION%%%%%%%%%%%%%%%%%%

\section{Related Work} \label{sec:related}

\noindent  
There has been a lot of recent work in utilizing Online Social Media (OSM) 
%for several fields of research namely opinion mining \cite{op-mine-2018}, event detection \cite{event-detect-2018}, early prediction of cyber-attacks \cite{cyber-bullying-2018}, temporal evolution of communities \cite{community-evolution-2018} and the like. 
%The ubiquity and pervasive nature of OSM has also inspired recent research on utilizing them 
to facilitate post-disaster relief operations -- see~\cite{social-media-emergency-survey,Nazer-disaster-osm-survey,disaster-info-mgmt-survey} for some recent surveys on this topic.  
For instance, there have been works on classifying situational and non-situational information~\cite{rudra-cikm-disaster,Rudra-tweb-2018}, location inferencing from social media posts during disasters~\cite{geotext,lingad,geolocalise-2018}, early detection of rumours from social media posts~\cite{Monda-2018}, emergency information diffusion on social media during crises~\cite{cindy-2018-ipm}, event detection~\cite{event-detect-2018}, extraction of event-specific informative tweets during disaster~\cite{lay-ipm} and so on.
Recent works have exemplified social media's ability to disseminate disaster information between institutional and non-institutional volunteers~\cite{Bush-abedin-18} and their use to reinforce the role of different stakeholders~\cite{Liu2018}.

There has also been research in designing information systems to aid emergency management~\cite{is-2010,dmm-2014}. A few of them have already been implemented in various countries and their efficacy has proven useful during disaster scenarios. AIDR~\cite{aidr-2014}, Ushahidi~\footnote{https://www.ushahidi.com/}, MapBox\footnote{https://osm-in.github.io/flood-map/chennai.html\#11/13.0000/80.2000} are some notable examples.
%Although in a nascent stage, we intend to incorporate this work as an information system.

Specifically, some prior studies attempted to identify OSM posts informing about need and availability of resources in a disaster situation. Varga {\it et al.}~\cite{varga-help-tweets} employed NLP based techniques to this problem on Japanese tweets posted during an earthquake in Japan.
%, most of the tweets in the dataset were in Japanese and techniques are difficult to reuse for English tweets. 
Purohit {\it et al.}~\cite{purohitFM} provided a set of $18$ regular expressions to identify specific types of need-tweets indicating donation requests for different resources, and tweets indicating the availability of resources to be donated.
There have also been attempts to develop tools that tap the capabilities of social media to identify need-requests and resource providers~\cite{seeker-supplier-2014,Nazer-request}.
In our prior works~\cite{BasuASONAM17,Khosla2017}, we perceived this problem of identifying need-tweets and availability-tweets as an Information Retrieval (search) task, where appropriate query-terms are selected to form queries corresponding to needs and availabilities.  
%Word embedding (Word2vec~\cite{Mikolov2013}) based methodology is employed to capture the semantics of need-tweets and availability-tweets and hence retrieved the tweets.  
Since traditional IR / NLP approaches often do not perform well over short, informally written microblogs, we used neural network-based IR models to identify need-tweets and availability-tweets. 
However, the works~\cite{BasuASONAM17,Khosla2017} did not make any attempt to match the need-tweets with appropriate availability-tweets.

%~\cite{BasuASONAM17,Khosla2017,purohitFM,varga-help-tweets}.
To our knowledge, only two prior works have attempted to address the problem of matching need-tweets and availability-tweets.
Purohit {\it et al.}~\cite{purohitFM} employed a supervised classification approach to match need-tweets and availability-tweets using regular expression matches and TF-IDF vectors as features. This method is difficult to reproduce, since the regular expressions used as features for matching and some other details are not specified in the paper. 
Hence, we implemented a simplified unsupervised version of the methodology in~\cite{purohitFM} as a baseline, using only the TF-IDF vectors.
Our recent work~\cite{matching-www18-poster} tried several methodologies for the matching problem, such as simple noun overlaps, and matching based on local word embeddings %using word2vec~\cite{Mikolov2013} 
and pre-trained word embeddings~\cite{pennington2014glove,imran2016lrec,twitter-paraphrase-TACL} over the text of the tweets. We consider all the methodologies in~\cite{matching-www18-poster} as baselines.

Thus, both the prior works~\cite{purohitFM,matching-www18-poster} considered the resources to be the sole factor for matching (or used the whole tweet text). 
Other important factors like source of the resource, quantity of the resource and geographical distance between the need and the availability were ignored.
These factors need to be considered in practice for post-disaster matching of needs and availabilities, and the present work takes the first step in this direction.
%Presently, as noted by \cite{imran-2018-survey}, there is a need for actionable insights rather than situational information for coordinating disaster-relief operations and we hope to achieve that with our work.

% Work done on logistics, resource transportation and storage (model it as an optimization problem)

%%%%%%%%%%%%%%%%%%%%% SECTION %%%%%%%%%%%%%%%%%%%%%%

%\section{\textcolor{blue}{Utility of the Proposed Methodologies in Developing a System for Aiding Post-disaster Relief Operations}} \label{sec:utility}
\section{Utility of the Proposed Methodologies in Developing a System for Aiding Post-disaster Relief Operations}\label{sec:utility}

\noindent One of the most prevailing challenges in post-disaster scenario is to deploy emergency resources such as food, water, shelter, medicines, etc. to the affected region in real-time. %However, due to the lack of first-hand information, the emergency resources are typically allocated based on adhoc estimates. 
During and in the aftermath of a disaster, emergency resources are often scarce. Thus, such resources must be mobilized and distributed in an optimized way, considering several factors like actual demand of a specific resource at a particular location, the geographical proximity between the locations of resource need and availability of resources, and many more.

However, in the aftermath of any disaster, disaster management authorities deal with severe uncertainties in allocating different resources, due to the lack of first-hand information regarding actual need and availability of specific resources in the disaster-affected regions.  
In this scenario, information posted on social microblogging sites like Twitter is increasingly being used for assisting post-disaster relief operations.  
The greatest advantage of social media platforms such as Twitter is that, due to the ubiquity of smartphones and mobile Internet, information can be obtained  from various stakeholders (including victims, relief workers) who are right at the site of the disaster, and that too in real-time.
However, the critical information that can help post-disaster relief operations is immersed within lot of conversational content (e.g., sympathy for victims); hence automated techniques are needed to extract such information quickly.
Therefore, disaster management authorities are in dire need of a system to identify tweets that inform about resource needs and availabilities (which we term as need-tweets and availability-tweets respectively) and also to automatically suggest how to match the needs with appropriate availabilities. To the best of our knowledge, such a system is not available till date.

As stated earlier, our work is the first attempt to develop methodologies to match need-tweets and availability-tweets considering important aspects such as geographical proximity (which were not considered in prior works~\cite{purohitFM,matching-www18-poster}). 
It should also be noted that our matching methodologies are unsupervised (unlike the supervised method in~\cite{purohitFM} which requires training), and hence can be readily applied to any ongoing disaster event after need-tweets and availability-tweets are identified, e.g., by the methodologies in~\cite{BasuASONAM17,Khosla2017}.
Hence we believe that the methodologies developed in the present work would be an important part of any system that assists post-disaster relief efforts.

%We hope that the acumen of our proposed system would assist disaster management authorities in deploying resources in an optimal way. Moreover, in future, we plan to develop more actionable matching methodology considering other important aspects of need-tweets and availability-tweets like the quantity of the resource, the trustworthiness of the source of resource availability etc.

%%%%%%%%%%%%%%%%%%%%%%%%%%%%%%%%
\section{Dataset Description} \label{sec:dataset}

\noindent 
For the present work, we re-use the datasets developed in our prior works~\cite{BasuASONAM17,Khosla2017,matching-www18-poster}. 
This section briefly describes the datasets. 
We collected tweets 
related to two major earthquake events --
(i)~the earthquake in Nepal in April 2015,\footnote{\url{https://en.wikipedia.org/wiki/April_2015_Nepal_earthquake}}
and 
(ii)~the earthquake in central Italy in August 2016.\footnote{\url{https://en.wikipedia.org/wiki/August_2016_Central_Italy_earthquake}}
For both events, we used the 
Twitter Search API\footnote{https://dev.twitter.com/rest/public/search} with the queries `nepal quake' and `italy quake' respectively,
to collect tweets that were posted during the days immediately following each event.
In total, about 100K tweets were collected for the Nepal earthquake,
and about 180K tweets for the Italy earthquake. 

Tweets frequently contain duplicates and near-duplicates due to 
retweets and re-posting of the same information by many users~\cite{Tao-duplicate-tweets}. 
Presence of duplicates can result in over-estimation of the performance 
of retrieval and matching methodologies.
%and can also create information overload for human 
%annotators while developing the gold standard~\cite{TREC-microblog-2015}.
Hence, we eliminated duplicate and near-duplicate tweets using a simplified version of the methodologies 
in~\cite{Tao-duplicate-tweets}.
Specifically, similarity of a pair of tweets was estimated by the Jaccard similarity
of the set (bag) of words contained in the two tweets (after ignoring stopwords, URLs and @user mentions). If two tweets were
found to be more similar than a threshold value of $0.8$, only one of the tweets was retained in the corpus.

After removing duplicates and near-duplicates, we obtained a set of 
{\it 50,068 tweets} for the Nepal earthquake dataset,
and {\it 70,487 tweets} for the Italy earthquake dataset.  
These sets were used for all experiments reported in this study.
For brevity, we will denote the two datasets as
{\it nepal-quake} and {\it italy-quake} respectively.

\vspace{2mm}
\noindent {\bf Identifying need-tweets and availability-tweets:} 
We re-use the need-tweets and availability-tweets that were identified in our prior works~\cite{BasuASONAM17,Khosla2017} through human annotators.
Three annotators were employed, each of whom is competent in English
and use Twitter regularly, but none of whom is an author of this paper. 
The annotators have significant prior experience in annotating disaster-related microblogs. 
For the nepal-quake dataset, the annotators found $499$ need-tweets in English, and $1,333$ availability-tweets in English. For the italy-quake
dataset, they found $177$ need-tweets and $233$ availability-tweets in English. More details of the dataset and identification of need-tweets and availability-tweets can be found in our prior works~\cite{BasuASONAM17,Khosla2017}.
The characteristics of the datasets are summarized in Table~\ref{tab:dataset}.

\begin{table}[tb]
\small
\centering
\begin{tabular}{|c|c|c|c|c|}
\hline
\textbf{Dataset} & \textbf{Total}	& \textbf{Distinct}	& \textbf{Need}	& \textbf{Availability} \\ 
 			   & \textbf{tweets}  & \textbf{tweets}     & \textbf{tweets}     & \textbf{tweets} \\
\hline
nepal-quake  & 100K & 50,068 & 499  & 1,333 \\
\hline
italy-quake  &  180K & 70,487 & 177  & 233  \\
\hline
\end{tabular}
\caption{{\bf Summary of datasets related to two disaster events.}}
\label{tab:dataset}
\end{table}

\vspace{2mm}
\noindent Note that, in the present work, we focus on need-tweets and availability-tweets in English language only. 
Our decision to focus on only English tweets was motivated by the observation that tweets referring to needs and availabilities are rarely written in local languages, probably because communication in English enables rapid communication between international agencies and the local population. 
For example, the need and availability tweets posted in Hindi during the Nepal earthquake were only $28$ and $34$ in number respectively, as compared to $499$ and $1333$ such tweets in English.

\vspace{2mm}
\noindent In the subsequent sections, we discuss methodologies to match each need-tweet (identified as described above) with suitable availability-tweets.

%%%%%%%%%%%%%%%%%%%%%%%%%%%%%%%%%%%%%%%%5
\section {Understanding need-tweets and availability-tweets} \label{sec:understanding}

\noindent 
We now attempt to understand the semantics of need-tweets and availability-tweets.
Based on our discussions with relief workers who regularly work in post-disaster situations\footnote{We discussed with relief workers from `Doctors For You' (\url{http://doctorsforyou.org/}) and SPADE (\url{http://www. spadeindia.org/}).}, we identified five particular information that are most needed to coordinate resource needs and availabilities:
\begin{compactitem}
\item {\bf Resource}: the item(s) that are needed or available, e.g., food, water, medicines, electricity.
\item {\bf Quantity}: how much of each resource is needed/available? This information is essential to understand the extent of the demand, so as to decide which demand to prioritize.
\item {\bf Location}: the geographical location where the resource is needed/available. Apart from knowing where the resource is needed, the distance between the need and availability is crucial to decide whether a particular resource-need can be satisfied by a particular available resource.
\item {\bf Source}: who needs the resource, or who is offering the resource (usually a person or an organization). During a major disaster, usually several government agencies / NGOs work in the same region; hence it is essential to know who needs a resource or is offering a resource, in order to coordinate activities of different relief organizations.
\item {\bf Contact}: whom to contact for information about the resource needed/available (e.g., phone number, emails). Contact information is valuable not only for establishing communication, but also for ensuring accountability.
\end{compactitem}
As advised by the relief workers, from a given need-tweet / availability-tweet, we attempt to extract the five types of information stated above.
In the rest of this section, we describe in detail how we extract the above information from need-tweets and availability-tweets after some initial preprocessing.

\vspace{2mm}
\noindent {\bf Tweet Preprocessing:}
We employed common pre-processing techniques on the tweet text to remove URLs (but not email ids), mentions, characters like brackets, `RT', and other non-ASCII characters like \#, \&, ellipses and Unicode characters corresponding to emojis. 
We also segmented CamelCase words and joint alphanumeric terms like `Nepal2015' into distinct terms (`Nepal' and `2015').
However, we did {\it not} perform case-folding on the text since we wanted to detect proper nouns. 
Likewise, we also abstained from stemming since the names of locations and sources might get altered.

\subsection{Extracting the resource(s)} \label{sub:identify-resources}

This section describes the proposed methodology for extracting resources from the tweet text. 
Note that our proposed methodology is very different from the resource extraction technique developed by Purohit {\it et al.}~\cite{purohitFM}.
Unlike the supervised technique of~\cite{purohitFM}, we adopt an unsupervised methodology that does not need any training data (which is expensive to gather in real-time during an ongoing disaster event). 
The second difference between our methodology and that of~\cite{purohitFM} is as follows. 
The methodology in~\cite{purohitFM} assigns a tweet a single label based on the type of the resource mentioned in the tweet text (e.g., cash, health, logistics -- see details below). 
However, imposing this `hard' class criterion, in which each tweet is related to a single resource type, is not viable, since a sizable fraction of tweets posted during a disaster can mention multiple resources. 
In fact, 75.8\% of the need-tweets  in the nepal-quake dataset refer to at least two distinct resources, e.g., 
``Mobile phones are not working, no electricity, no water in \#Thamel \#Nepalquake''.
Thus, in our methodology, an individual tweet can be mapped to more than one resource-type, as detailed below.

From a given tweet, we first identify a set of words / phrases which may potentially be resources. Next, we match these potential resources with a pre-compiled list of resources that are frequently needed in disaster situations. We now describe these steps.

\begin{figure}[tb]
\centering
\includegraphics[height=7.5cm, width=0.7\columnwidth]{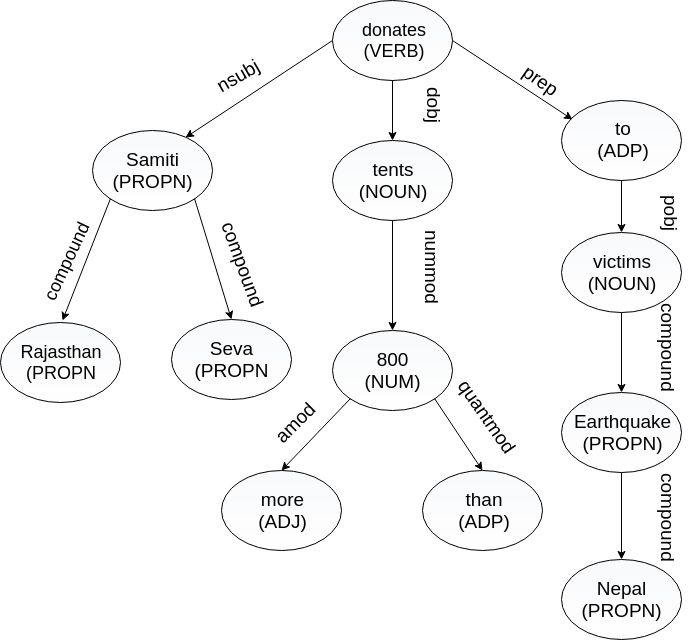}
\caption{{\bf Dependency tree generated for the tweet `Rajasthan Seva Samiti donates more than 800 tents to Nepal Earthquake victims'. The arcs represent the dependencies between the words and their corresponding parts-of-speech are enclosed within brackets.}}
\label{fig:Resource_dependency}
\end{figure}

\vspace{2mm}

\noindent {\bf Algorithm to extract potential resources:}
The algorithm for extracting potential resources comprises of the following steps.
\begin{itemize}
\item We perform dependency parsing on the preprocessed tweet-text to obtain a dependency tree (using the dependency parser of SpaCy (\url{https://spacy.io/}).
Figure~\ref{fig:Resource_dependency} illustrates the dependency tree for a sample tweet. 
%the tweet ``Rajasthan Seva Samiti donates more than 800 tents to Nepal Earthquake victims''. 

\item From prior works~\cite{BasuASONAM17,Khosla2017}, we obtain a set of words that are commonly used to signify needs (e.g., `need', `require') and availabilities (e.g., `available', `distribute', `send', `donate') of resources in a disaster scenario. We refer to these two sets of words as {\it need-words} and {\it availability-words}.
From the dependency tree of a tweet, we identify the ROOT-word and any need-word or availability-word, and include them in a list of {\it head words}. 

\item We then observe the dependencies of the child-nodes of these head words. Specifically, we consider those dependencies where the child is an object (direct, indirect), subject (nominal, clausal) or conjunct of one of the head words, and treat those child-nodes as potential resources.
Since resources are likely to be nouns, we only consider those child-nodes as potential resources, which have been tagged as a common/proper noun by a Part-Of-Speech (POS) Tagger. 
%This restriction is because almost all resources are nouns. 

\item If the dependency of any child-node corresponds to a prepositional modifier, clause modifier (adverbial, relative)  complement (adjectival, clausal, open clausal) or punctuation,
we include it in the list of head words and recursively observe the dependencies of their corresponding child-nodes.
\end{itemize}

\noindent  
For the tweet shown in Figure~\ref{fig:Resource_dependency}, the algorithm returns `tents', `Samiti' and `victims' as potential resources because they are respectively the dobj (direct object), pobj (prepositional object) and nsubj (nominal subject) of the availability word `donate'.

Using the above algorithm, we effectively obtain a set of nouns / noun phrases that are potential resources, since they are closely connected with the need-words or availability-words. 
%Next, we match these potential resources with a pre-defined list of resources that are frequently needed in disaster situations.

\vspace{2mm}
\noindent {\bf Matching potential resources with a pre-compiled list of resources:}
Having extracted a set of potential resources from the text of a tweet, we now check for semantic similarity of the extracted words with a comprehensive list of resources that are commonly needed in disaster situations.
We describe how we compile the comprehensive list of resources.

The United Nations Office for the Coordination of Humanitarian Affairs (UNOCHA)
has identified five broad classes of resources that are frequently needed in disaster scenarios\footnote{\url{https://vosocc.unocha.org/getFile.aspx?file=att36103_h4t800.pdf}, Section 2.3.2} -- 
(i)~cash (monetary contributions to the government / non-government organizations working in the affected areas), 
(ii)~health (field hospitals, medical kits, medicines, etc.),
(iii)~logistics (transportation modules (helicopters, 4x4 vehicles), forklifts, storage facilities, etc.),
(iv)~shelter (tents, tarpaulin, rebuilding kits, etc.), and
(v)~water and sanitation (hygiene kits, water purification systems, water distribution kits, portable latrines, etc.).
Prior works in this domain~\cite{purohitFM} have followed similar categorizations of resources.
Following the UNOCHA guidelines, we construct a comprehensive list of resources from the five classes, as follows.

%The aforementioned 7 categories are similar to the 6 mutually exclusive classes as noted in \cite{purohitFM}, namely money, clothing, food, medical supplies (including blood), shelter and volunteer work. This forms the basis of the supervised classification framework wherein the authors categorized each of the need and avail tweets into one of these 6 classes.

%In a fashion similar to the guidelines laid down by OCHA  we subdivide the resources into five major categories i.e. medical, food, shelter, cash and logistics.
\noindent \underline{Food class:}
The food class comprises of foodstuffs vital to disaster response planning, as noted in~\cite{food15}, and other stockpiled rations which are essential emergency supplies \footnote{\url{https://www.realsimple.com/food-recipes/shopping-storing/emergency-foods}}.

\noindent \underline{Health class:} We include 
(i)~the medical items recommended by the World Health Organization (WHO) as listed in~\cite{ghanamed}, 
(ii)~a generic essential emergency equipment list \footnote{http://www.who.int/surgery/publications/s15982e.pdf}, and 
(iii)~sanitation requirements adhered to by Center for Disease Control and Prevention (CDC) and WHO.\footnote{\url{https://www.sswm.info/water-nutrient-cycle/wastewater-treatment/hardwares/sanitation-emergencies/immediate-and-short-term-emergency-sanitation} and \url{https://www.cdc.gov/healthywater/global/sanitation/sanitation-emergency-response.html}}

\noindent \underline{Shelter class:}
These resources are enlisted from a number of lists\footnote{\url{https://emergency.unhcr.org/entry/60043}, and \url{https://preparednessmama.com/72-hour-kits-clothing-and-shelter/}}. 
as well as from~\cite{shelter15} that highlights the minimum standards of shelter and settlement for the disaster-afflicted victims. 

\noindent \underline{Logistics class:}
%Logistics and volunteer work form the crucial backbone of coordinated disaster relief efforts. 
We enlist multi-dimensional types of logistic resources, ranging from trained personnel like doctors, transport, power and electricity communication, storage, equipments etc.\footnote{\url{http://crisis.med.uoa.gr/elibrary/15.pdf}}  

\noindent \underline{Cash class:} we consider terms that are common synonyms of `cash', such as `funds', `donations' and `money'. 

\vspace{1mm}
\noindent Table~\ref{tab:resource_types} states some examples of the resources enlisted in each class by the process described above. 

\begin{table}[tb]
\centering
\small
    \begin{tabular}{|c|c|} 
    \hline
    Category & Examples \\ \hline
    Health & blood, anesthetic, antibiotics, latrines, tissue paper, sanitary napkins, soap bars\\
    \hline
    Shelter& tents, rope, tarpaulins, sheets, blankets, clothes, shelter kit, jackets, boots, gloves, camp \\
    \hline
    Food & cereal, bottled water, canned food, utensils, fuel, dried fruits, biscuits, vegetables,\\
    \hline
    Logistics & electricity, storage, doctors, army, power, helicopters, communication, volunteers, \\
    \hline
    Cash & Funds, money, supplies, donations, stock\\    
    \hline
\end{tabular}
\caption{{\bf Some examples of resources of the five classes}}
\label{tab:resource_types}
\end{table}
 
%We employ the POS Tagger and Dependency Parser of SpaCy \footnote{https://spacy.io/} to extract the resources. 
We match the set of potential resources extracted from the text of a tweet (using the algorithm described earlier) with this list of resources commonly needed in disaster situations. 
If the extracted word/phrase appears in this list of commonly-needed resources, we include it directly. 
Otherwise, we check the semantic similarity of the extracted word/phrase with the commonly-needed resources using the well-known Wordnet ontology~\cite{wordnet}. 
We recognize the extracted word/phrase as a resource if its semantic similarity with any of the resources in the list is higher than $0.8$. 

For the example tweet shown in Figure~\ref{fig:Resource_dependency}, recall that three terms were extracted as potential resources -- `tents', `Samiti' and `victims'. 
Out of these, only `tents' is recognized as a viable resource by the methodology described above.   

% Specifically, for each potential resource, we check its semantic similarity with the resources in the list using the well-known Wordnet ontology~\cite{wordnet}.
%the inbuilt Wordnet module of NLTK~\citet{wordnet}. 

\subsection{Extracting geographical locations}  \label{sub:identify-location}

\begin{figure}[tb]
	\centering
		\includegraphics[height=11cm, width=0.7\columnwidth]{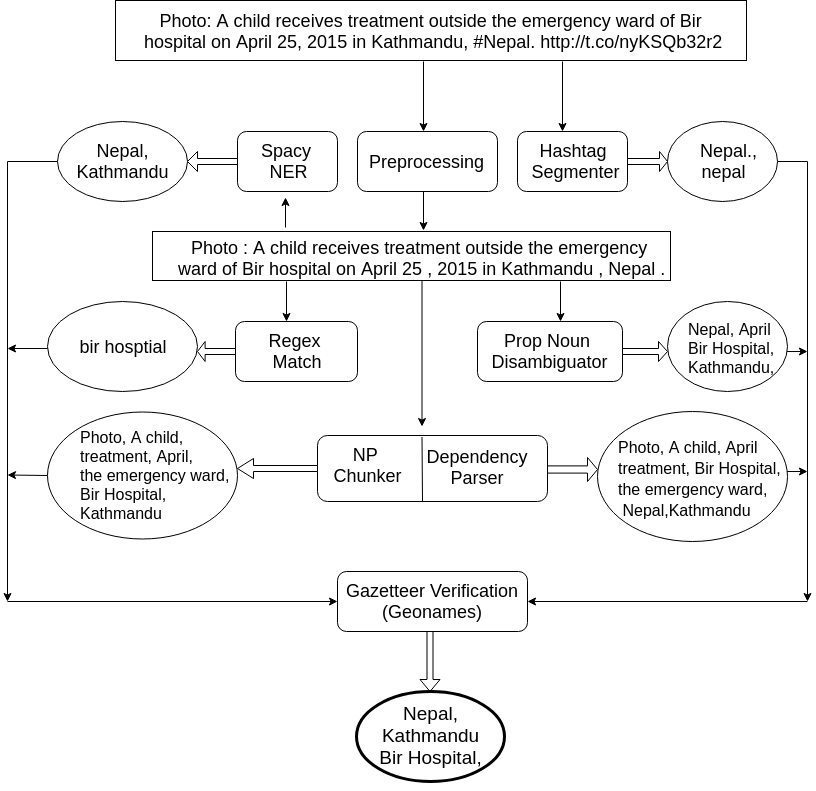}
		\caption{{\bf Flowchart depicting the functioning of our algorithm for extracting geographical locations, on a sample tweet ``Photo: A child receives treatment outside the emergency ward of Bir hospital on April 25, 2015 in Kathmandu, \#Nepal [url]''. The locations identified are `Nepal', `Kathmandu' and `Bir hospital'.}}
		\label{fig:Flowchart_location}
\end{figure}

Only a very small fraction of tweets are geo-located, i.e., have geographical coordinates associated with them~\cite{social-media-emergency-survey}. Hence we attempt to infer geographical locations from the text of the tweets, using
an unsupervised methodology
that is an improved version of the method in our prior work~\cite{savitr-smerp18}.
We now describe each of the steps involved.
Figure~\ref{fig:Flowchart_location} diagrammatically shows each step of the methodology for a sample tweet.

\vspace{2mm}
\noindent {\bf Hashtag Segmentation:}
Tweets posted during disasters often have location names embedded in hashtags, e.g., \#NepalQuake (Nepal earthquake), \#chennaiFloods (floods in the Indian city of Chennai), \#HoustonStrong (cyclone in Houston, USA).
%Due to the peculiar style of coining hashtags, it becomes imperative to break them into meaningful words.
Hence, we attempt to segment hashtags into meaningful words using a statistical word segmentation based algorithm~\cite{Peter}, similar to the approach in~\cite{Malmasi} and~\cite{DBLP:journals/corr/abs-1708-03105}. 
Along with the word segments, 
we also retain the original hashtag, to ensure that we do not lose meaningful remote locations simply because they are uncommon and hence undetected by statistical methods.

\vspace{2mm}
\noindent {\bf Disambiguating Proper Nouns from Parse Trees:}
Since location names are generally
composed of one or more proper nouns, we apply a POS (Parts-of-Speech) tagger 
on the preprocessed tweet-text to extract the relevant POS tags. 
Let the $i^{th}$ word of the tweet be $w_i$ and its corresponding POS tag be denoted by $T_{i}$. If $T_{i}$ happens to be a proper noun, we append the words trailing $w_i$, provided their POS tags correspond to proper nouns, or adjectives or delimiters (conjunctions (`and', `or') or punctuations (`;', `)' ). 
Additionally, we utilize a list of common suffixes of location names to recognize locations.
The suffix list -- a part of which is shown in Table~\ref{tab:example-list} -- comprises different naming conventions for
landforms\footnote{https://en.wikipedia.org/wiki/List\_of\_landforms}, 
roads\footnote{https://wiki.waze.com/wiki/India/Editing/Roads} \footnote{http://www.haringey.gov.uk}, 
buildings\footnote{https://en.wikipedia.org/wiki/List\_of\_building\_types} and towns.

If $w_i$ is followed by a noun in this suffix list or has high Jaro-Winkler similarity~\cite{jaro-winkler} ($\geq$ 0.75) with one in the suffix list, we consider it to be a viable location. 
% Since Out-of-Vocabulary (OOV) words are common in Twitter, we also consider those words which have a high Jaro-Winkler similarity with the words in the suffix list.

If the word immediately preceding $w_i$ is a preposition that usually precedes a place or location, such as `at', `in', `from', `to', `near', etc., or directions like `north', `eastern', etc (see Table~\ref{tab:example-list}), we consider it a viable location as well.  
%We then split the stream of words obtained via the delimiters.
Thus, we attempt to infer from the text proper nouns which conform to locations from their syntactic structure.

%\subsubsection{Regex matches}
\vspace{2mm}
\noindent {\bf Regex matches:}
We perform a regular expression match to identify phrases that contain any of the suffixes or prefixes mentioned in Table~\ref{tab:example-list}. 
%Likewise, the prefix list specifies words that prepend a location, like `southern' California. 

\begin{table}[tb]
\centering
\small
	\begin{tabular}{|c|c|}
	\hline
	Type & Common Examples\\ \hline
	
	Landforms & doab, lake, steam, river, island, valley, mountain, hill \\
	Roads & street, st, boulevard, junction, lane, rd, avenue, bridge\\
	Buildings & hospital, school, shrine, cinema,villa, temple, mosque,  \\
	Towns & city, district, village, gram, place,town, nagar,  \\
	Directions & south, eastern, NW, SE, west, western, north east,\\ \hline
   
	\end{tabular}
  	\caption{{\bf Examples of suffixes/prefixes of locations}}
	\label{tab:example-list}
\end{table}

\vspace{2mm}
\noindent {\bf Dependency Parsing of needs and availabilities:}
Here we leverage the dependency graph generated during resource extraction (as described in Section~\ref{sub:identify-resources}). % to identify viable locations. 
Our objective is to identify locations where resources are needed or available. To this end, we identify nouns, adjectives and proper nouns which are at a short {\it dependency distance} of 3-4 units from the need-words and availability-words (which were explained in Section~\ref{sub:identify-resources}).
The dependency distance is measured by the number of links connecting the words in the dependency graph of the tweet text. 
A short dependency distance implies the word is more intimately related to the need-word or availability word. 
This phase is useful as it helps to identify potential locations which do not conform with the aforementioned heuristics. 

We also extract the noun phrases from the dependency graph, 
and then use an NER tagger to identify named entities which might conform to locations (in a way similar to~\cite{Malmasi,lingad,gelernter}). 
%We have leveraged the NER Tagger provided by SpaCy as opposed to the more commonly available NER tools like Stanford NER~\cite{StanfordNER}, 
%Twitter NLP~\cite{TwitterNLP1}, Open NLP~\cite{OpenNLP}, due to the faster execution time of the former as noted in \todo {cite SAVITR}.

\vspace{2mm}
\noindent {\bf Gazetteer Verification:}
Having extracted a list of potential phrases and words, only those that correspond to real-world locations are retained using a gazetteer. 
The gazetteer verification step is essential to filter out redundant nouns / noun phrases obtained via dependency parsing and the proper nouns identified by the proper noun disambiguator as depicted in Figure \ref{fig:Flowchart_location}. Additionally, the gazetteers also return the geo-spatial coordinates which enables us to calculate the distance between the locations where a resource is needed and where it is available. 

The choice of the gazetteer depends upon the granularity and precision of locations that we intend to identify.
We explored the use of two gazetteers, namely Geonames (\url{http://www.geonames.org/}) and Open Street Map (\url{http://geocoder.readthedocs.io/providers/OpenStreetMap.html}). 
%and incorporated both into the algorithm. 
We observed that Geonames is comparatively faster but lacks fine-grained information of buildings and roads, as compared to Open Street Map. Consequently, if a viable location is appended by a word conforming with naming conventions for buildings or roads (like `Bir Hospital' in Figure~\ref{fig:Flowchart_location}), we detect it using the Open Street Map gazetteer.

\subsection{Extracting Quantities, Sources, and Contact} \label{sub:quan-source-contact}

\noindent 
Now we describe how we extract the other three fields, namely quantities, sources and contacts. 

\vspace{4mm}
\noindent {\bf Quantities:}
Having extracted the list of possible resources by the method described in Section~\ref{sub:identify-resources}, we check whether the resources are preceded by some numeric token in the original tweet text. 
A token is considered numeric if its orthographic notation corresponds to a real number (say `$1000$') or it semantically embodies a real number (say `thousand'). 
For instance, in Figure~\ref{fig:Resource_dependency}, which corresponds to the tweet ``Rajasthan Seva Samiti donates more than 800 tents to Nepal Earthquake victims'', `800' is a numeric token which precedes the resource `tents' and hence is identified as a quantity with respect to the resource `tents'. 
Note that if a tweet contains multiple resources, there can be a quantity associated with each resource.

\vspace{4mm}
\noindent {\bf Sources:}
In a manner similar to extracting resources, we perform dependency parsing on the processed tweet text. 
We then identify from the dependency tree, subjects of the corresponding head words (described in Section~\ref{sub:identify-resources}), which are also tagged as {\it proper nouns}. 
We also identify proper nouns from the tweet text corresponding to locations, persons and organizations. 
We filter out the words which have been identified as resources or locations (using the aforementioned algorithms for resource extraction and location inferencing) and identify the rest as viable sources.
Note that the source may comprise of one or more words. Hence, if a particular proper noun is detected as a source, we check whether it is compounded with other proper nouns; if so, we consider all those consecutive proper nouns as the source.

For instance, the resource extraction algorithm identified `tents' as a  resource from the tweet text ``Rajasthan Seva Samiti donates more than 800 tents to Nepal Earthquake victims.'', as illustrated in Figure~\ref{fig:Resource_dependency}.  Since `Samiti' is tagged as a proper noun and is a nominal subject (nsubj) to the verb `donate', it was identified as a source. 
Since `Samiti' is compounded to other proper nouns, we recognize `Rajasthan Seva Samiti' as a source for this tweet. 

\vspace{4mm}
\noindent {\bf Contacts:} We extract two types of contact information - email ids and phone numbers - from the pre-processed tweet-text using common regular expressions as shown in Table~\ref{tab:contact-list}.
%We extract three vital pieces of information that can be deemed as  contact information, namely email, phone number and url links from the preprocessed tweet text. 
%These are identified by common regular expression matches as shown in Table~\ref{tab:contact-list}
%The contact information is obtained from the preprocessed text by simple regex matches. 

\begin{table}[tb]
\centering
\small
    \begin{tabular}{|p{0.1\linewidth}|p{0.7\linewidth}|}
	\hline
	Type & Regex\\ \hline
	PhoneNum &
    \verb|([+]?[0]?[1-9][0-9\s]*[-]?[0-9\s]+)|
    \\
    Email & \verb|[a-zA-Z0-9]?[a-zA-Z0-9_.]+[@][a-zA-Z0-9_.]+[.](com|$|$net$|$edu$|$in$|$org$|$\verb|en)| \\
 
\hline
	\end{tabular}
  	\caption{Regex-es for extracting contact information}
	\label{tab:contact-list}
\end{table}

\subsection{Baseline methodology for parsing the need-tweets and availability-tweets} \label{sub:baseline-extraction}

\noindent 
Earlier in this section, we described our proposed methodology to parse need-tweets and availability-tweets. One can question whether sophisticated tools like deep parsers are actually needed for this task, or whether simpler tools such as POS taggers and NER tools are sufficient. This question specifically arises since, for many non-English resource-poor languages, tools such as deep parsers are not available; hence it needs to be seen what level of performance can be achieved without using such sophisticated tools. 
In this section, we consider a simple baseline methodology for understanding need-tweets and availability-tweets, that utilize only POS taggers and NER tools. The five fields are identified as follows.

\noindent 
\textbf{Resources:} We use a POS tagger to identify common nouns, proper nouns, and noun phrases in the tweet text, which forms our set of potential resources. We then check for semantic similarity of the extracted words with the comprehensive list of resources described in Section~\ref{sub:identify-resources}, and extract the matching words as resources. \\
\textbf{Locations:} We employ a NER tagger to identify named entities that conform to locations, namely GPE, LOC, and FACILITY tags reported by the NER tagger. The extracted locations are then verified using the gazetteer described in Section~\ref{sub:identify-location}. \\
\textbf{Sources:} The geographical entities which are not identified by the gazetteer and the named entities that correspond to PERSON, NORP (Nationalities or religious or political groups) or ORG are identified as potential sources.\\
\textbf{Quantities and Contacts} are identified in the same way as described in Section~\ref{sub:quan-source-contact}. A numeric token preceding an identified resource is considered a quantity, and regular expressions are used to identify contacts.

\vspace{2mm}
\noindent In this section, we described two methodologies (one proposed and the other a simple baseline) for extracting the five fields -- resource, location, quantity, contact, source -- from need-tweets and availability-tweets. 
Note that a particular tweet may report about more than one resources. Also, a tweet may be both a need-tweet and an availability-tweet, mentioning about the need of one resource and availability of another. In such cases, our extraction methodologies attempt to identify the other fields for each resource.

The extraction methodologies will be evaluated later in Section~\ref{sec:evaluation}.
We now proceed to discuss methodologies for matching needs and availabilities.

%%%%%%%%%%%%%%%%%%%%%%

\section{Matching needs and availabilities}  \label{sec:matching-methods}

\noindent 
In this section, we discuss several methodologies for matching need-tweets and availability-tweets. The methods include both baseline methods from prior works, and proposed methods based on the information extracted as discussed in the previous section.
For a particular need-tweet, each of the methods identifies a list of matching availability-tweets, ranked
in decreasing order of how closely they match the need-tweet.

\subsection{Matching based only on resources: proposed methods and baselines} \label{sub:resource-matching-methods}

\noindent 
As stated earlier, the methodologies in prior works~\cite{purohitFM,matching-www18-poster} attempted matching only based on the resources mentioned in the tweet-text. 
To compare with these methodologies, we propose some methodologies that are based only on the resources.
These methods are important since many of the need-tweets and availability-tweets do not mention other information such as locations, and matching such tweets must be on the basis of the resources.

\vspace{2mm}
\noindent \underline{{\bf Proposed methodologies}}\\
We discuss methods to match need-tweets and availability-tweets based on the resources identified using the methodology in Section~\ref{sub:identify-resources}.

\vspace{1mm}
\noindent {\bf (P1) Matching based on common resources:}
We check for common resources mentioned in the need-tweets and availability-tweets. 
Specifically, for a given need-tweet, we compute the fraction of 
the resources extracted from this tweet, that are also contained in an availability-tweet.
For the given need-tweet, availability-tweets are ranked in the decreasing order of the fraction
of common resources (ties resolved arbitrarily).

\vspace{1mm}
\noindent {\bf (P2) Matching based on embeddings of resources:}
The method (P1) can identify a match only if a need-tweet and an availability-tweet mentions exactly the same resource. 
But, as shown in Table~\ref{tab:need-available-examples}, often different tweets mention semantically similar resources but using different terms (e.g., need of `shelter' and availability of `tents'). 
To identify such matches, we use word embeddings that can capture the semantic context of terms.

\vspace{1mm}
\noindent {\bf (P2a) Using local embeddings:}
Here we train Word2vec~\cite{Mikolov2013} on the set of tweets from where the need-tweets and availability-tweets have been
identified. 
Word2vec gives a term-vector for each term in the set of tweets, where the term-vector is expected to capture the semantic context of the term.
We construct a {\it resource-vector} for each tweet, by averaging the term-vectors of the resources extracted from the tweet (note that Word2vec vectors are additive~\cite{Mikolov2013}). 
The match between an availability-tweet  and a need-tweet is given by the {\it cosine similarity of their resource-vectors}. 
For a given need-tweet, availability-tweets are ranked in the decreasing order of this cosine similarity value (ties resolved arbitrarily).

\vspace{1mm}
\noindent {\bf (P2b) Using pre-trained embeddings:}
A limitation of using local embeddings is that a large set of tweets about the current event is needed to learn the local word embeddings. 
At times of a disaster, it is necessary to start matching needs and availabilities even before a sizable number of tweets about the current event have been collected.
To this end, we explore the use of pre-trained embeddings.
We use Word2vec embeddings pre-trained over tweets related to several disaster or crisis events~\cite{imran2016lrec}, which can be expected to capture well the semantics of terms used during disaster events.
As in methodology (P2a), we compute a resource-vector
for each tweet, and then compute the cosine similarity between the resource-vectors of need-tweets and availability-tweets.
For a particular need-tweet, matching availability-tweets are ranked in decreasing order of this cosine similarity.

\vspace{2mm}
\noindent \underline{{\bf Baseline methodologies from prior works}}\\
Following the works~\cite{purohitFM,matching-www18-poster}, we used the following baseline approaches.

\noindent {\bf (B1) Matching based on common nouns~\cite{matching-www18-poster}:} A part-of-speech tagger is applied to identify {\it nouns} from need-tweets and availability-tweets,
since nouns are likely to indicate the specific resources.\footnote{We use the Twitter-specific part-of-speech tagger developed by Gimpel {\it et al.}~\cite{postag-2012};
available at \url{http://www.cs.cmu.edu/~ark/TweetNLP/}.} 
For a given need-tweet, availability-tweets are ranked in the decreasing order of the fraction
of common nouns in the need-tweet that are also contained in the availability-tweet (ties resolved arbitrarily).

\noindent {\bf (B2) TF-IDF based matching:} 
As stated in Section~\ref{sec:related}, the supervised matching methodology of Purohit {\it et al.}~\cite{purohitFM} could not be reproduced since some of the details are not public.
Hence we use a simplified version of the methodology in~\cite{purohitFM}, where we compute TF-IDF vectors for the tweets (where term frequencies and inverse document frequencies are measured as specified in~\cite{purohitFM}).
We compute the match between a need-tweet and an availability-tweet by the cosine similarity of their TF-IDF vectors.
For a given need-tweet, matching availability-tweets are ranked in decreasing order of this cosine similarity.

\noindent {\bf (B3) Matching based on local word embeddings~\cite{matching-www18-poster}:} 
Word2vec~\cite{Mikolov2013} is trained on the set of tweets from where the need-tweets and availability-tweets have been
identified. A tweet-vector is constructed for each tweet, by averaging the Word2vec term-vectors of all the terms contained in the tweet. 
The match between an availability-tweet and a need-tweet is computed as the {\it cosine similarity of their tweet-vectors}. 
%(similar to the methodology (P2a)). 
This method is similar to the proposed methodology (P2a), the difference being that while (P2a) specifically considers the resources extracted from the tweet, this baseline methodology considers all words in a tweet.
%For a given need-tweet, availability-tweets are ranked in the decreasing order of this cosine similarity value (similar to the proposed methodology (P2a)). 

%between the tweet-vectors of the availability-tweets and that of the need-tweet.
%The intuition behind our proposed approach is to match the overall semantic context of the need-tweet and the availability-tweet, which
%is likely to be able to match tweets even if exact word-level matching is not possible. 

\noindent {\bf (B4) Matching based on pre-trained word embeddings~\cite{matching-www18-poster}:} 
Three types of pre-trained word embeddings were used in~\cite{matching-www18-poster}:

\noindent (B4a) {\it Pre-trained GloVe embeddings~\cite{pennington2014glove}}, which are pre-trained on two billion tweets from the Twitter 1\% random sample~\cite{pennington2014glove}.

\noindent (B4b) {\it Pre-trained embeddings for detecting paraphrases in tweets~\cite{twitter-paraphrase-TACL}}: These embeddings are designed to detect paraphrases in tweets, i.e., tweets which are lexically different but semantically similar~\cite{twitter-paraphrase-TACL}. Hence, these embeddings can potentially identify needs and availabilities that are written differently, but have similar semantics.

\noindent  (B4c) {\it Word2vec embeddings pre-trained over tweets related to several disaster or crisis events~\cite{imran2016lrec}}, which are also used in the proposed methodology (P2b).

\noindent For all the pre-trained embeddings,  a tweet-vector is computed
for each tweet, and then matching is performed based on the cosine similarity between the tweet-vectors of need-tweets and availability-tweets, as done in~\cite{matching-www18-poster}.

\subsection{Proposed methodology for matching based on resources and geographical proximity} \label{sub:resource-location-matching-methods}

\noindent 
All the methods described above consider only the resources for the purpose of matching. 
However, matching needs and availabilities in practice in a post-disaster scenario needs to consider other aspects as well, such as the geographical proximity of the need and the availability of the resource. 
Hence, we now propose a matching methodology that considers both the resources mentioned in the tweets, as well as their geographical locations (as identified using the methodology in Section~\ref{sub:identify-location}).
This particular methodology can be used only when the location is identified for both a need-tweet and an availability-tweet.

Note that there has been some prior work that infers locations from need-tweets and availability-tweets~\cite{purohit-location}, but none of the prior works have used the geographical information during matching.

From a given need-tweet and an availability-tweet, we extract the resources and the locations. Then we compute two similarity scores between the tweets, one based on the resources and the other based on the locations mentioned in the tweets:

\noindent \underline{\it Resource similarity score:}
For matching the resources, we use the methodology (P2b) -- cosine similarity of the resource-vectors computed using pre-trained word embeddings -- since this is the method that performs the best matching of resources, as compared to the other methods (see detailed experiments in the next section). This score measures the similarity of the resources mentioned in the two tweets.

\noindent \underline{\it Proximity score:}
For checking the geographical proximity, we consider the geographical coordinates of the locations extracted from the two tweets and compute the distance between the two locations. 
We normalize this distance using a bounding box around the country where the disaster has occurred (e.g., a bounding box around Nepal or Italy). Thus we get a
proximity-score that captures the geographical proximity between the location of the need and the location of the availability.

\noindent \underline{\it Matching score:}
The final matching-score between the need-tweet and the availability-tweet is computed as a linear combination of the resource-similarity score and the proximity-score. 
For now, we give the same weightage of $0.5$ to the two scores.

\noindent 
For a given need-tweet, availability-tweets are ranked in decreasing order of this matching-score, with ties resolved arbitrarily.

%The methodologies described above are evaluated in the next section.

%%%%%%%%%%%%% SECTION %%%%%%%%%%%%%%%%%

\section{Evaluation of methodologies} \label{sec:evaluation}

\noindent 
We now evaluate the performance of methodologies proposed earlier in the paper -- (i)~the methodology for extracting the five types of information from need-tweets and availability-tweets, and 
(ii)~the matching methodologies described in Section~\ref{sec:matching-methods}.

\subsection{Evaluating information extraction methodologies}

\noindent 
Here we evaluate the proposed methodology and the simple baseline methodology for extraction of the five types of information from need-tweets / availability-tweets (described in Section~\ref{sec:understanding}).

\vspace{2mm}
\noindent \textbf{Evaluation methodology and measures:} 
We randomly selected $50$ need-tweets and $50$ availability-tweets from each of the two datasets (nepal-quake and italy-quake), i.e., $200$ tweets in total.
Then we asked three human annotators (the same who had identified the need-tweets and availability-tweets) to check the five fields -- resources, quantity, location, source and contact -- extracted from each of the selected tweets by the proposed algorithm, and judge if the extracted information was correct.
There was almost unanimous agreement between the judgments of all the annotators (for over 95\% cases); majority voting was considered in the few cases where there was no unanimous agreement.
Based on the annotators' judgment, we compute the following measures, separately for each of the five fields:

\noindent (i) Precision: what fraction of the information extracted by the algorithm is correct.

\noindent (ii) Recall: every field is not mentioned in every tweet. The Recall measures what fraction of each field that was mentioned in the tweets could be correctly extracted by the algorithm.

\noindent (iii) F-score: the harmonic mean of the Precision and Recall.

It should be noted that, all need-tweets / availability-tweets do {\it not} contain all the five fields. We asked our set of annotators to report the statistics of need-tweets and availability-tweets having precise information about the five important aspects. According to our annotators, for the need-tweets of nepal-quake dataset, the percentages of tweets containing the five fields are as follows -- resource (93\%), location (8\%), source (45\%), contact (2\%)  and quantity (4\%). For the availability-tweets of the nepal-quake dataset, the corresponding percentages are resource (92\%), location(9\%), source (61\%),  contact (4\%), quantity(2\%). For the italy-quake dataset, the percentages are lower.
Hence, while evaluation of the information extraction methodologies, we consider an extraction to be correct if it matches the opinion of our annotators in both cases (whether the corresponding information is present in the tweet or not).

\begin{table}
\small
\centering
\begin{tabular}{|p{1.3cm}||c|c|c||c|c|c|}
\hline
\textbf{Field} & \textbf{Prec} & \textbf{Recall} & \textbf{F-Score} & \textbf{Prec} & \textbf{Recall} & \textbf{F-Score}\\			 
\hline
 & \multicolumn{3}{|c||}{{\bf nepal-quake}} & \multicolumn{3}{|c|}{{\bf italy-quake}}\\
\hline
Resources & 0.9591   & 0.8246  & 0.8868  &0.8333  &0.8064  & 0.8196 \\ \hline
Location  & 0.9354  & 0.8529  & 0.8922   & 0.8571 &0.7500  & 0.8000   \\ \hline
Quantity  & 0.6923  & 0.9000  & 0.7826     & -- & -- & -- \\ \hline
Source    & 0.6154  & 0.5333 & 0.5714   &0.5000 &0.6667  & 0.5714  \\ \hline
Contact   & 1.0000   &1.0000   &1.0000 & 1.0000   &1.0000   &1.0000   \\ \hline
\end{tabular}
\caption{{\bf Evaluation of the {\it proposed} methodology for extracting the five fields from need-tweets for the two datasets. Quantity information was not present in any of the need-tweets of italy-quake dataset; hence left blank.}}
\label{tab:eval-extraction-needs}
\end{table}

\vspace{2mm}
\noindent \textbf{Evaluation of the proposed methodology:}
Table~\ref{tab:eval-extraction-needs} and Table~\ref{tab:eval-extraction-avails} illustrate the performance of our proposed methodology for  extracting the five fields from need-tweets and availability-tweets for the two datasets.  
The tables state the precision, recall and F-score for each of the five fields, that is achieved by the proposed methodology for need-tweets (Table~\ref{tab:eval-extraction-needs}) and availability-tweets (Table~\ref{tab:eval-extraction-avails}).
The Quantity information was not present in any of the need-tweets of italy-quake dataset; hence the corresponding entries are left blank in Table~\ref{tab:eval-extraction-needs}. 
For both datasets, our proposed algorithm performs well in extracting the  Resource, Location and Contact fields -- for each of these three fields, the F-Score value is higher than $0.8$. 
In extracting Source and Quantity information, our methodology also performed reasonably well, as evident from the result shown in Table~\ref{tab:eval-extraction-needs}and Table~\ref{tab:eval-extraction-avails}.

\begin{table}
\small
\centering
\begin{tabular}{|p{1.3cm}||c|c|c||c|c|c|}
\hline
\textbf{Field} & \textbf{Prec} & \textbf{Recall} & \textbf{F-Score} & \textbf{Prec} & \textbf{Recall} & \textbf{F-Score}\\			 
\hline
 & \multicolumn{3}{|c||}{{\bf nepal-quake}} & \multicolumn{3}{|c|}{{\bf italy-quake}}\\
\hline
Resources  & 0.8400 & 0.8873   & 0.8630   &0.8387  &0.8214  & 0.8299 \\ \hline
Location   & 0.8695 & 0.8333  & 0.8510   &0.8000  & 1.0000 & 0.8888  \\ \hline
Quantity   & 0.7826 & 0.8846   & 0.8305   &0.8571  &0.8571  & 0.8571  \\ \hline
Source     & 0.6800   & 0.6800  & 0.6800   &0.8333  &0.6250  & 0.7142  \\ \hline
Contact    & 1.0000  & 1.0000   & 1.0000 & 1.0000  & 1.0000   & 1.0000   \\ \hline
\end{tabular}
\caption{{\bf Evaluation of the {\it proposed} methodology for extracting the five fields from  availability-tweets for the two datasets.}}
\label{tab:eval-extraction-avails}
\end{table}

Table~\ref{tab:field-extraction-examples} shows some example of need-tweets and availability-tweets and the corresponding extracted fields. 
Wrongly extracted fields are shown in red and italic font. 
We see that the methodology identifies `KTM' (acronym of Kathmandu) as Location accurately. For several tweets containing quantity information for multiple resources (e.g., `Pakistan Army is sending 2000 meals, 200 tents, medicine, 600 blankets 2 Nepal'), all the resources and quantities are accurately identified by our methodology. 
However, the methodology failed to identify the quantity `four tons' (in the fourth tweet shown in Table~\ref{tab:field-extraction-examples}), probably because the word `tons' was not considered as a valid numeric identifier. 

Also, the table shows few examples where the correct resource could not be detected, primarily because the need / availability was mentioned in a covert way. For instance, the phrase `people are staying outside' was used to express need for shelter, and `please remove passwords from working wi-fi connections' was used to express need for connectivity (see Table~\ref{tab:field-extraction-examples}).
In such cases, no need-word was detected (as explained in Section~\ref{sub:identify-resources}), hence the corresponding resources could not be identified.

\begin{table*}
 \footnotesize
 \centering
 \resizebox{1\textwidth}{!}{%
 \begin{tabular}{|p{0.37\textwidth} |p{0.15\textwidth}|p{0.10\textwidth}|p{0.155\textwidth}|p{0.13\textwidth}|p{0.09\textwidth}|}
 \hline
  \textbf{Tweet text (excerpts)} &\textbf{Resource} & \textbf{Location} &\textbf{Quantity} &\textbf{Source} & \textbf{Contact}\\
  \hline
   Pakistan Army is sending 2000 meals, 200 tents, medicine, 600 blankets 2 Nepal
   &medicine, meals, tents, blankets 
   &nepal
   &tents-200, blankets-600, meals-2000 & Pakistan Army & \\
   \hline

   Blood donations apparently shortage of certain types. Please call Dr Manita at 98XXX-XXXXX &blood & & & Dr Manita &98XXX-XXXXX\\\hline
   Punjabi Shikh community in KTM, providing free food at Kupandole Gurudwara (near Bagmati bridge) & food & ktm, bagmati& & \textit{\textcolor{red}{Kupandole Gurudwara}} & \\
   \hline
        Lists of help provided to Nepal in Earthquake reliefs by India. 95 Tons food packets, 94 Tons Water & Water, reliefs, food packets & nepal &  & India & \\
   \hline
   electricity is available at Shankhamul area, people are staying outside Kathmandu& electricity & kathmandu & & \textit{\textcolor{red}{people}}, \textit{\textcolor{red}{Shankhamul}}&\\
   \hline
    Modi govt sent four Ton relief material, Team of doctors, 10 x NDRF, JCBs, food, water, medicines to Nepal & water, food, medicines,  \textit{\textcolor{red}{Ton material}}, doctors & Nepal &  & govt, NDRF & \\
    \hline
 Earthquake Collection efforts taking place in Florence today from 10 am - 10 pm (asking for clothes, shoes, water) & clothes, shoes, water, \textit{\textcolor{red}{pm}} & florence & \textit{\textcolor{red}{pm-10}} & &\\
   \hline
  Italy Red Cross deploys Relief units and Rescue dogs , 20 ambulances on their way & ambulances, Relief, Rescue dogs & italy & ambulances-20 &&\\ 
   \hline
   If in the area affected by earthquake in Italy please remove passwords from working wi-fi connections and for emergencies call 800 XXX XXX & &italy & & \textit{\textcolor{red}{emergencies}} & 800 XXX XXX\\
   \hline
 \end{tabular}
 }
 \caption{{\bf Examples of information extracted from need-tweets and availability-tweets by the proposed methodology. The items that are wrongly extracted are marked in red italics.}}
\label{tab:field-extraction-examples}
\end{table*}

\vspace{2mm}
\noindent
\textbf{Evaluation of the baseline methodology:}
Next we evaluate the baseline methodology that uses only a POS tagger and a NER tagger for extracting the five fields, without any deep parsing (described in Section~\ref{sub:baseline-extraction}).
For this evaluation, we use the same 50 need-tweets and 50 availability-tweets from the two datasets, that were used to evaluate the proposed methodology.
Since both the proposed and baseline methods use the same approach for identifying contact and quantity, we do not evaluate the performance of the baseline method for these two fields.
We now report the performance of the simple methodology in identifying the resource, source, and locations.

We observe the baseline methodology performs reasonably well in identifying resources. The methodology is able to identify almost 77\% of the resources correctly. However, in identifying location and source, the methodology performed very poorly (less than 20\% precision). 
The reason for such poor performance is probably because the simple NER tagger could not identify many geographical entities mentioned in the tweets, which thus could not be verified using the gazetteer.  
Additionally, the baseline methodology often wrongly identifies locations as sources, since geographical entities which are not identified by the gazetteer as locations are recognized as potential sources by this methodology.  
For example, in the tweet ``\textit{Everybody is sheltering on open sky in Kathmandu tonight, having very poor access to food, water and latrine}'', the term `Kathmandu' is wrongly identified as a source. 
Similarly, in the tweet ``\textit{Search and rescue dogs and 20 ambulances on the ground in Perugia following earthquake volunteers from on the scene}'', resources such as ambulances are identified correctly, but `Perigua' is wrongly identified as source instead of location.

Thus, we conclude that the use of deep parsing in our proposed methodology is justified, and not using the deep parser would lead to significantly poor performance in extracting some of the fields.

\subsection{Evaluating matching methodologies}

\noindent 
Now we evaluate the methodologies for matching need-tweets with availability-tweets, which were described in Section~\ref{sec:matching-methods}.

\vspace{2mm}
\noindent \textbf{Evaluation methodology and measures:} 
For a given need-tweet, each matching methodology produces a ranked list of availability-tweets.
For each need-tweet, we used the methodology being evaluated to retrieve $5$ top-matching availability-tweets.
Thus, if there are $n$ need-tweets (in the gold standard), we obtain $5 \times n$ need-availability pairs.
We then asked three human annotators (the same who identified the need-tweets and availability-tweets, as described in Section~\ref{sec:dataset}) to check each need-availability pair, and judge whether the match is correct.

\noindent \underline{When is a need-availability match considered correct?} 
For the methodologies that match only based on resources (described in Section~\ref{sub:resource-matching-methods}), an availability-tweet $t_a$ is considered a correct match for a need-tweet $t_n$ if $t_a$ mentions availability of at least one resource whose need is stated in $t_n$. 
The annotators were asked to consider the semantics of the resources, so that availability of `tents'  would be considered to correctly match the need for `shelter'.

For the proposed methodology that matches based on both resources and location (described in Section~\ref{sub:resource-location-matching-methods}), the annotators were asked to consider two conditions. First, the condition stated above was used for judging the matching of resources. As a second condition, 
%was applied to judge the geographical proximity of the need and the availability. 
the annotators were asked to check the locations mentioned in the need-tweet $t_n$ and the availability-tweet $t_a$, and to look up the distance between the two locations using Google Maps. 
$t_a$ is considered a good match for $t_n$ if the road-distance between the two locations is less than 100 kilometers, or if the travel-time between the locations is less than 4 hours, as informed by Google Maps (these thresholds were suggested to us by the relief workers with whom we discussed).

We found very high agreement between the judgements of all the annotators (for over 96\% cases); majority voting was considered in the few cases where there was no unanimous agreement.
We then use the following evaluation measures to judge the performance of a matching methodology.

\noindent \underline{(i) Precision of matching:} 
The precision of a matching methodology is the fraction of pairs that are 
matched correctly by the methodology. 
In Information Retrieval terminology, we are computing {\it Precision@5.}

\noindent \underline{(ii) Recall of matching:} 
The recall of a particular methodology is the fraction of all need-tweets for which the methodology 
is able to identify at least one correct matching (based on the judgment of the annotators over the top $5$ availability-tweets retrieved by the methodology).

%It is important to note that, for a given dataset, it might not be possible to match some of the  need-tweets
%because there are no corresponding availability-tweets in the dataset (e.g., a need-tweet is about a resource which was not at all available).
%Hence, we first needed to identify the subset of need-tweets which had at least one matching availability-tweet (in the gold standard).
%We engaged our human annotators to manually check the gold standard sets,
%and identify the subset of need-tweets which had a matching availability-tweet.
%We considered a need-tweet to have a matching availability-tweet if a majority of the annotators judged so. 

\noindent \underline{(iii) F-score of matching:} Finally, the F-score of a matching methodology is the harmonic mean
of the precision and recall. 
%We use the F-score as the primary measure for comparing between two matching methodologies. 

\vspace{2mm}
\noindent \textbf{Results:}
Table~\ref{tab:matching-nepal} compares the performance of different matching methodologies for the nepal-quake dataset, while
Table~\ref{tab:matching-italy} compares that for italy-quake dataset. 
Both tables state the performances of the baseline methodologies, the proposed methods considering resources only, and the proposed methodology considering resources and location.

\begin{table}[tb]
\small
%\scriptsize
%\footnotesize
\centering
\begin{tabular}{|p{10cm}|c|c|c|}
\hline
\textbf{Methodology for matching} & \textbf{ Precision} & \textbf{Recall}	 & \textbf{F-Score}\\			 
\hline
\multicolumn{4}{|c|}{{\bf Baseline methodologies}}\\
\hline
(B1) Common noun overlap & 0.5350 & 0.8323 & 0.6768 \\ 
\hline
(B2) TF-IDF based matching & 0.7741 & 0.9032 & 0.8337  \\ 
\hline
%(B2) TF-IDF based matching & 0.5455 &0.9010  & 0.6796  \\   % from evaluation of an old method
%\hline
(B3) Local Word2vec embeddings & 0.7676 & 0.9535 & 0.8505\\
\hline
(B4a) Pre-trained Glove embeddings of tweet text & 0.1903 & 0.5676 & 0.2850 \\ 
\hline
(B4b) Pre-trained paraphrase embeddings~\cite{twitter-paraphrase-TACL} of tweet-text  & 0.2622 & 0.6768 & 0.3780 \\ 
\hline
(B4c) Pre-trained crises-specific embeddings~\cite{imran2016lrec} of tweet-text  & 0.2756 & 0.7151 & 0.3979 \\ 
\hline
\hline
\multicolumn{4}{|c|}{{\bf Proposed methodologies based on resources}}\\
\hline
(P1) Common resources & 0.7900   & 0.9167   & 0.8486   \\
\hline
(P2a) Local embeddings of resources & 0.7533   &  0.8833  & 0.8131   \\
\hline
(P2b) Pre-trained embeddings of resources &  {\bf 0.8364}  & {\bf 0.9616}    & {\bf 0.8946} \\
\hline
\hline
\multicolumn{4}{|c|}{{\bf Proposed methodology based on resources \& location}}\\
\hline
Pre-trained embeddings or resources + geographical proximity &0.7312  &0.9062  & 0.8093 \\
\hline
\end{tabular}
\caption{{\bf Evaluation of matching need-tweets and availability-tweets for the nepal-quake dataset.}}
\label{tab:matching-nepal}
\end{table}

\begin{table}[tb]
\small
%\footnotesize
\centering
\begin{tabular}{|p{10cm}|c|c|c|}
\hline
\textbf{Methodology for matching} & \textbf{ Precision} & \textbf{Recall}	 & \textbf{F-Score}\\			 
\hline
\multicolumn{4}{|c|}{{\bf Baseline methodologies}}\\
\hline
(B1) Common noun overlap & 0.6727 & 0.9242 & 0.7786 \\ 
\hline
(B2) TF-IDF based matching & 0.6762 &0.7679  & 0.7191 \\ 
\hline
%(B2) TF-IDF based matching &0.3061  & 0.7424 &0.4335  \\   % from evaluation of an old method
%\hline
(B3) Local Word2vec embeddings  of tweet-text &   0.4000 & 0.7727 & 0.5271 \\
\hline
(B4a) Pre-trained Glove embeddings  of tweet-text &  0.2848 & 0.6970 & 0.4043 \\ 
\hline
(B4b) Pre-trained paraphrase embeddings~\cite{twitter-paraphrase-TACL}  of tweet-text  & 0.4000 & 0.7727 & 0.5271 \\ 
\hline
(B4c) Pre-trained crises-specific  embeddings~\cite{imran2016lrec}  of tweet-text  & 0.3030 & 0.7121 & 0.4251 \\ 
\hline
\hline
\multicolumn{4}{|c|}{{\bf Proposed methodologies based on resources}}\\
\hline
(P1) Common resources & 0.8000   &  {\bf 0.9545}   & 0.8704  \\
\hline
(P2a) Local embeddings of resources & 0.6848   &  0.9090   & 0.7811  \\
\hline
(P2b) Pre-trained embeddings of resources & {\bf 0.8090}  & {\bf 0.9545}    & {\bf 0.8757} \\
\hline
\hline
\multicolumn{4}{|c|}{{\bf Proposed methodology based on resources \& location}}\\
\hline
Pre-trained embeddings or resources + geographical proximity & 0.5058 & 0.9411 & 0.6580 \\

\hline
\end{tabular}
\caption{{\bf Evaluation of matching need-tweets and availability-tweets for the italy-quake dataset.}}
\label{tab:matching-italy}
\end{table}

\noindent \underline{Matching based on resources only:}
Among the methodologies that consider resources only, we observe that our proposed methods achieve substantially better performance than the baselines from prior works. 
Specifically, the method based on the disaster-specific pre-trained embeddings of resources (P2b) performs best for both the datasets. 
This good performance is probably due to the use of disaster-specific word embeddings which can capture the semantics of resources.
%For the italy-quake dataset, performance of the methodology based on the common resources (P1) is quite close to the performance of P2b.  
%Table~\ref{tab:need-available-examples-new} shows some examples of correctly and incorrectly matched need-tweets and availability-tweets (considering the resource only) identified by the best performing methodology (P2b). 

Table~\ref{tab:need-available-examples} showed some examples of correctly matched need-tweets and availability-tweets (considering the resource only) identified by the best performing methodology (P2b).
For each pair of need-tweet and availability-tweet, the common resource is shown in bold. 
%It is apparent from the Table~\ref{tab:need-available-examples-new} that some pairs are easy to match since they refer to exactly the same resource (e.g., `water' or `medicine'),
It is evident that the methodology is able to match pairs that are tricky to match because they use different terms to refer to the same resource (e.g., need of `power' and availability of `electricity', need of `bread and roof' and availability of `shelter and food'). 
The methodology also correctly matches tweets containing references to multiple resources. 
However, we noticed that the method fails to match some rare resources (e.g., `antibiotics', `swabs', `electricity') probably because these terms are not present in the vocabulary of prior events on which the model is pre-trained.

\vspace{2mm}
\noindent \underline{Matching based on resources and location:}
Table~\ref{tab:matching-nepal} and Table~\ref{tab:matching-italy} also
report the performance of our methodology that considers both resource and geographical proximity of the need and the availability of the resource for matching. 
Note that this methodology is evaluated only for those need-tweets that contained a location. 
The method achieves Recall higher than $0.9$ in both datasets, but the precision is higher for nepal-quake ($0.7312$) than for italy-quake ($0.5058$).
%performs significantly well in case of the nepal-quake dataset.  However, for the Italy-quake dataset, the precision is quite low, 
This difference in precision is probably because the italy-quake dataset contains very limited references to locations, and the few locations mentioned are far apart from each other, so that at times the threshold of geographical proximity considered in our evaluation is not satisfied.

\begin{table}
\footnotesize
 %\scriptsize
\centering
\begin{tabular}{|p{0.4\columnwidth} | p{0.4\columnwidth} |p{0.13\columnwidth}|}
\hline
\textbf{Need-tweet (excerpts)} & \textbf{Availability-tweet (excerpts)}& \textbf{Location proximity}\\
\hline
\multicolumn{3}{|c|}{{\bf Correctly Matched Examples}}\\
\hline
\textbf{\textit{TU teaching hospital}}. \#Earthquake injured are getting treatment on open space. \textbf{Tent}s are urgent need  & Distribution of \textbf{tents} to the earthquake victims at \textbf{\textit{Tistung}} by fellow Rotarians & 59.6 km
\\ \hline
@MEAcontrolroom Indians stranded in Nepal's \textbf{\textit{Sundhara}} relief camp need help. Without \textbf{food} or water & Distributing \textbf{food} in \textbf{\textit{Tudikhel}} Relief work for Nepal \#earthquake by \#ArtofLiving & 2.1 km\\
\hline
Italy earthquake urgent \textbf{blood} donations needed \textbf{\textit{\#Rieti}}  & People gathered this morning in \textbf{\textit{Rome}} to donate \textbf{blood} for the injured \#earthquake & 80 km\\
\hline
%American military must send URGENT \textbf{RESCUE TEAMS} to free those trapped under rubble in \textbf{\textit{Perugia's}} Earthquake & \textbf{Maltese volunteers} head out to help earthquake-stricken \textbf{\textit{Umbria}} [url] & 87.7 km\\
%\hline
\hline
\multicolumn{3}{|c|}{{\bf Incorrectly Matched Examples (proximity more than threshold)}}\\
\hline
\#Earthquake Collection efforts taking place in \textbf{\textit{\#Florence}} today from 10am-10pm (asking for \textbf{clothes, shoes, water}) & Photo: \textbf{Water} collected for earthquake victims, in \textbf{\textit{Rieti}}, Italy - Rieti Life via Facebook via @Breaking & 267.6 km\\
\hline
Need for \textbf{helicopters} in \textbf{\textit{\#Langtang!}} Dutch travelers need to be rescued & The latest on Nepal quake: \textbf{Helicopters} ferry the injured to \textbf{\textit{Gorkha}} & 878 km\\
\hline
Rainfall start in \textbf{\textit{bhaktapur}} nepal, create another problem for the people in camps,  clean \textbf{drinking water} is needed & Sahara Club Sends  \textbf{Bottled Water}, Noodles To \textbf{\textit{Gorkha}} For Earthquake Victims & 152.6 km\\
\hline
 \end{tabular}
\caption{{\bf Examples of need-tweets and matching availability-tweets, as identified by the proposed methodology considering resources and locations. The common resource for each pair is shown in bold. The locations for each pair is shown in bold and italic.}}
 \label{tab:need-available-examples-location}
\end{table}

Table~\ref{tab:need-available-examples-location} shows some examples of correctly and incorrectly matched pairs of need-tweets and availability-tweets, as matched by the proposed methodology. %considering both resource matching and location proximity. 
For each pair, the table also shows the road-distance between the locations of the need and the availability, as obtained from Google Maps.
%For each pair of need-tweet and availability-tweet, the common resource is in bold and the location in bold italic. 
It can be seen that the geographical proximity of the need and the availability of the resource is accurately captured by our methodology and matched accordingly.  
Some of the matchings were judged incorrect due to our considered proximity threshold of 100 kilometers, as shown in Table~\ref{tab:need-available-examples-location}; however, for most of these cases as well, the resource matching was correct.
%The method matches location name in hashtags ( e.g., `\#Reiti'). 
%The methodology also identifies the reference of location indirectly from organization names ( e.g ., for both  `TU teaching hospital'  or `ArtOfLiving Nepal Centre' the location, is `Kathmandu') and matched accordingly.
	
\section{Reusing the methods for other disaster events} \label{sec:reuse}

\noindent 
We now study the reusability of our proposed methodologies on a completely different type of disaster event. To this end, we experimented with a dataset of tweets posted during floods in 2015 in the Indian city of Chennai\footnote{https://en.wikipedia.org/wiki/2015\_South\_Indian\_floods}.  
We used the queries `chennai' and `flood' to collect tweets using Twitter Search API. We collected 215K English tweets and removed duplicates and near duplicates and finally a set of  66K distinct English tweets was obtained. The same set of annotators were employed  for identification of need-tweets and availability-tweets (as described in Section~\ref{sec:dataset}). 
The annotators identified $4,516$ need-tweets and $6,583$  availability-tweets for the chennai-floods dataset.

We applied our best performing methodology (as observed for the previous two datasets), i.e., the method
based on the disaster-specific pre-trained embeddings of resources (P2b), on the tweets related to this new disaster event. We evaluate the methodology using the same approach as described in the previous section.

We also experimented with the matching methodology that considers both pre-trained embeddings of resources and geographical proximity of the need and the availability of the resource. Since the chennai-floods disaster occurred within a single city (as compared to country-wide events such as nepal-quake and italy-quake), in the case of chennai-floods, we considered the threshold of geographical proximity between a resource-need and a resourc-availability to be only {\it 20 km} (as opposed to 100 km for the country-wide events). In other words, we consider a need-tweet to be correctly matched by an availability-tweet (both containing geographical locations) if the distance between their locations is less than 20 km according to Google Maps.

Table~\ref{tab:matching-chennai} reports the performances of the methodologies on the chennai-floods dataset.
We find that the methods perform reasonably well in case of chennai-floods dataset. The matching considering resources only achieved a precision of $0.84$ and a recall of $0.87$.
The matching methodology that consider both resources and locations also achieve reasonably high precision~($0.76$) and recall~($0.96$).  
Table~\ref{tab:need-available-examples-location-chennai} shows some examples of correctly and incorrectly matched pairs of need-tweets and availability-tweets by the proposed methodology (based on geographical proximity) on chennai-floods dataset.

These results demonstrate that our proposed methodologies can be readily used for different types of disaster events.

\begin{table}[tb]
%\small
%\scriptsize
\footnotesize
\centering
\begin{tabular}{|p{8cm}|c|c|c|}
\hline
\textbf{Methodology for matching} & \textbf{ Precision} & \textbf{Recall}	 & \textbf{F-Score}\\			\hline
\multicolumn{4}{|c|}{{\bf Proposed methodology based on resources}}\\
\hline
(P2b) Pre-trained embeddings of resources &  {\bf 0.8434}  & 0.8696    & {\bf 0.8562 } \\
\hline
\hline
\multicolumn{4}{|c|}{{\bf Proposed methodology based on resources \& location}}\\
\hline
Pre-trained embeddings or resources + geographical proximity &0.7565  &0.9565  & 0.8448 \\
\hline
\end{tabular}
\caption{{\bf Evaluation of matching need-tweets and availability-tweets for the chennai-floods dataset.}}
\label{tab:matching-chennai}
\end{table}

\begin{table}
\footnotesize
 %\scriptsize
\centering
\begin{tabular}{|p{0.4\columnwidth} | p{0.4\columnwidth} |p{0.13\columnwidth}|}
\hline
\textbf{Need-tweet (excerpts)} & \textbf{Availability-tweet (excerpts)}& \textbf{Location proximity}\\
\hline
\multicolumn{3}{|c|}{{\bf Correctly Matched Examples}}\\
\hline
6 womens need \textbf{food} at \textbf{\textit{hostel No 1, 1st Street,Jai Nagar, Valasaravakkam, Chennai - 600087, Near Holy Cross School}} & \textbf{Food} for 300+ people available in \textbf{\textit{KK Nagar, Chennai}}. Ready now. For those who need  & 3.8 km\\
\hline
@MEAcontrolroom B-ve \textbf{blood} needed 4 Surgery, \textbf{\textit{Apollo Greams Road }}Call [mobile no] & 6 Units O+ve \textbf{blood} at\textbf{\textit{ Vijaya Hospital Vadapilini}} Pls Call [mobile no] @BloodDonorsIn @iCanSaveLife  & 7.6  km\\
\hline
Anna i am from \textbf{\textit{MKB Nagar}}. Please concentrate on North Chennai. More than 1000 people are homeless without \textbf{food} and \textbf{shelter} & Cheer NGO can provide \textbf{shelter} and \textbf{food} for 50 ppls @ \textbf{\textit{ Taylors Road, Kilpauk, Chennai - 10}} & 8.2 km\\
\hline
%American military must send URGENT \textbf{RESCUE TEAMS} to free those trapped under rubble in \textbf{\textit{Perugia's}} Earthquake & \textbf{Maltese volunteers} head out to help earthquake-stricken \textbf{\textit{Umbria}} [url] & 87.7 km\\
%\hline
\hline
\multicolumn{3}{|c|}{{\bf Incorrectly Matched Examples (proximity more than threshold)}}\\
\hline
\textbf{Blankets} n \textbf{bedsheets} are needed in \textbf{\textit{Manali, Redhills, Tiruvotriur n ponneri}} areas & \textbf{Blankets}  were distributed by Swayamsevaks in \textbf{\textit{Cuddalore, Chennai}}& 210 km\\
\hline
 \end{tabular}
\caption{{\bf Examples of need-tweets and matching availability-tweets, as identified by the proposed methodology considering resources and locations. The common resource for each pair is shown in bold. The locations for each pair is shown in bold and italic.}}
\label{tab:need-available-examples-location-chennai}
\end{table}

\section{Conclusion and Future Directions}\label{sec:conclusion}

\noindent 
We addressed the problem of matching resource needs and availabilities from microblogs posted during a disaster event,
which is important for effective coordination of post-disaster relief operations. 
We proposed an algorithm to extract important information from need-tweets and availability-tweets, including the resources, location, the quantity, sources and contacts. 
We also proposed methodologies to match needs with availabilities considering only resources (which out-performed baselines from prior works), as well as the first methodology to match considering both resources and locations.
%We tried different methods for a simplified version of the matching problem,  
%, where an availability-tweet is assumed to match a need-tweet if they mention the same resource., 
In this concluding section, we discuss some future extensions of the work.
\begin{itemize}

\item The actual problem of matching needs and availabilities of resources in a disaster situation is a complex one. 
Given that new needs and availabilities appear over time at different locations, a spatio-temporal resource optimization problem can actually be formulated for the matching.
For now, we have leveraged only the geographical proximity between the need and available resources. Incorporating the temporal features would require the solution to be an online one. 
Our proposed methodologies in this paper would help in formulating and solving the practical spatio-temporal resource optimization problem, which we plan to pursue as future work.

\item From the examples of need-tweets and availability-tweets given in various Tables, it is evident that all the need-tweets and availability-tweets are not equally {\it actionable}. What constitutes actionable information is not fully defined; still, one can assume that a tweet which mentions the location of the need/availability is more actionable than another tweet which does not mention any location.
Similarly, the tweets having precise information of source, contact, and quantity along with the resource name are definitely more actionable. 
As stated earlier, only a small fraction of need-tweets and availability-tweets contain the fields such as locations, sources, etc.
%We have already reported the percentages of need-tweets and availability-tweets having specific
%Information about at least one of the five important aspects of need-tweets and availability-tweet. 
In this particular study, we focused on matching need-tweets and availability-tweets considering both resource similarity and geographical proximity. The work can be extended to give higher priority to more actionable tweets for matching. In addition, the actionable tweets can possibly be better identified by actual relief workers. So, in future, we look to actively involve relief workers to this end.

\item A practical matching scheme should incorporate priority of various resource-needs. 
For instance, lack of water for 24 hours is possibly a more pressing need than lack of food for 6 hours. Thus, the available resources should be utilized to first address the prioritized needs. 
Another practical extension of this work, is the necessity to differentiate between the two different types of availability of  resources -- potential availability (which indicates availability in future) and present availability.

\item Finally, the present study only considered tweets in English. During a disaster event in a multilingual region, tweets in different non-English languages are also posted. Hence a practical extension would be to extend the methodologies for non-English languages. 
Such extension would be especially challenging for resource-poor languages, e.g., for which no deep parser is available. In this context, we found that our simple parser (with POS tagger and NER tool, but no deep parsing) achieved a reasonable accuracy  (77\%)  in identifying resources present in a tweet. Thus, in future, we can use this simple tagger to identify resource reference present in non-English tweets and hence could extend our matching methodology to compute resource similarity to match non-English tweets as well.

\end{itemize}

\section*{Acknowledgement} 

\noindent The authors thank the anonymous reviewers for their valuable comments which helped to improve the paper.
This research was partially supported by a grant from SRIC, IIT Kharagpur.
R. Dutt and S. Ghosh also acknowledge support from Microsoft Research India -- part of the work was done at the 2017 MSR India Summer Workshop on Artificial Social Intelligence. Especially, the authors acknowledge Dr. Monojit Choudhury of MSR India for useful discussions in the initial stages of the work.

\section*{References}

%\bibliography{mybibfile}

\end{document}